# *In silico* modeling of the rheological properties of covalently crosslinked collagen triple helices

*David A. Head\*†, Giuseppe Tronci‡[a], Stephen J. Russell‡ and David J. Wood[a]*

†School of Computing, University of Leeds, UK

‡Nonwovens Research Group, School of Design, University of Leeds, UK

[a]Biomaterials and Tissue Engineering Research Group, School of Dentistry, St. James's University Hospital, University of Leeds, UK

**Table of Contents Graphic**

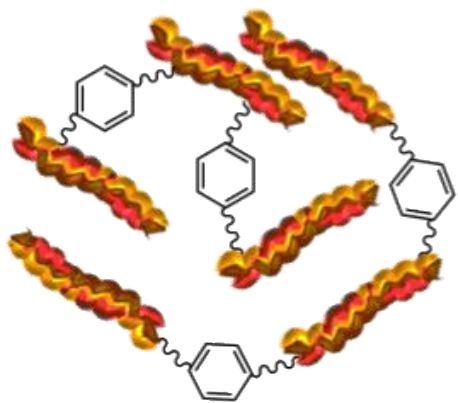 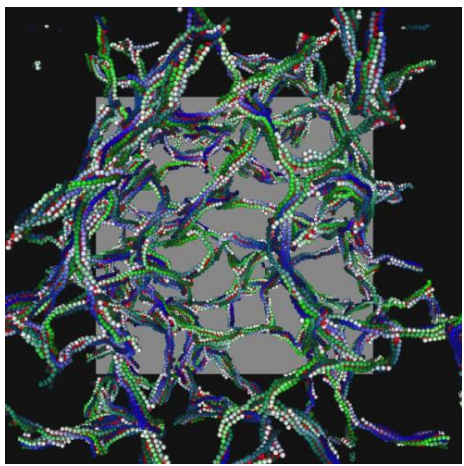

**Abstract**

Biomimetic hydrogels based on natural polymers are a promising class of biomaterial, mimicking the natural extra-cellular matrix of biological tissues and providing cues for cell attachment, proliferation and differentiation. With a view to providing an upstream method to guide subsequent experimental design, the aim of this study was to introduce a mathematical model that described the rheological properties of a hydrogel system based on covalently crosslinked collagen triple helices. In light of their organization, such gels exhibit limited collagen bundling that cannot be described by existing fibril network models. The model presented here treats collagen triple helices as discrete semi-flexible polymers, permits full access to metrics for network microstructure, and should provide a comprehensive understanding of the parameter space associated with the development of such multi-functional materials. Triple helical hydrogel networks were experimentally obtained *via* reaction of type I collagen with both aromatic and aliphatic diacids. The complex modulus $G^*$ was found from rheological testing in linear shear and quantitatively compared to model predictions. *In silico* data from the computational model successfully described the experimental trends in hydrogel storage modulus with either (i) the concentration of collagen triple helices during crosslinking reaction or (ii) the type of crosslinking segment introduced in resulting hydrogel networks. This approach may pave the way to a step change in the rational design of biomimetic triple helical collagen systems with controlled multi-functionality.

## 1. Introduction

Hydrogels consist of a hydrated, highly porous percolating network spanning macroscopic dimensions, in which the hydrophilic or amphiphilic building blocks remain insoluble due to crosslinking agents[1,2] or reversible physical interactions[3,4]. When formed from biomolecules they can share structural and functional similarities to natural extracellular matrices, making them attractive candidates for medical and dental applications[1]. A promising class of hydrogel is that formed from type I collagen. This protein, the most prevalent in the human body, is 300nm in length, 1.5nm in diameter, and its monomeric form consists of a triple helix of predominately polyproline chains[5,6]. In biological tissues, collagen triple helices are folded into fibrils, fibers and fascicles, which are stabilized *via* covalent crosslinks. Collagen hydrogels based on covalently crosslinked collagen triple helices[7] have recently been reported as a promising class of biomimetic systems in light of the controlled structure-property relationships[8], material format tunability[9] and large swelling ratio and compressive modulus[2]. As such, they can find application as biomimetic scaffolds for regenerative medicine[10,11], as well as non-woven wound dressings[12,13] and vascular grafts[14,15].

The mechanical properties of hydrogels are often crucial to their intended function[1,2,16,17]. Scaffolds must be able to support embedded cells while maintaining structural integrity when mechanically challenged, and must transmit forces from the environment to the cells as this affects many aspects of developing cell physiology, including differentiation[18-20]. Cues can also be dynamic, *i.e.* time or frequency-dependent, in native environments subject to habitual loading[21-23]. However, without rational design principles, controlling the mechanical properties of collagen hydrogels remains challenging, potentially resulting in compromised biomaterials with suboptimal efficacy[2,24].

The development of bespoke hydrogel scaffolds tailored to specific functions would be greatly accelerated if their mechanical properties could be reliably predicted without recourse to large scale *in vitro* assays. For collagen fibrils with diameters in the range 100-500nm resulting from the aggregation of collagen triple helices under physiological conditions[25], models of crosslinked elastic filament networks have been developed that compare favorably to *in vitro* experiments[7,26-28]. Whilst fibrillogenesis can be induced *ex vivo*, fibrillar collagen is barely tunable since selective functionalization and crosslinking is difficult to accomplish due to the unique collagen hierarchy[29,30]. When employed as a biomaterial, collagen systems built from covalently crosslinked triple helices can be tailored *via* systematic variation of triple helix concentration, degree of functionalization and network architecture. Here, bundling of the individual collagen molecules is hindered[1,2,30], resulting in networks whose filaments are just a few nanometers in thickness. For such materials, where thermal fluctuations will play an important role[25,31,32], no validated model currently exists. In addition, the bulk mechanics derives from a combination of the properties of the individual macromolecules, and the network microstructure, *e.g.* the pore size, connectivity *etc.*, which is not known *a priori* for non-physiological collagen networks[16,33]. Such complexity can, however, be represented in computational models that explicitly evolve networks in time, in contrast to the procedurally-generated networks currently used in models for collagen fibril networks[7,26-28]. In this way, an additional *in silico* model layer is introduced upstream from traditional *in vitro* experiments, enabling the identification of candidate regions of parameter space for a reduced number of experiments to explore.

The aim of this investigation was to construct and evaluate a mathematical model able to describe the rheological behavior of hydrogels formed by covalently crosslinking type-I collagen. Our thermal model is intended to represent hydrogel networks built from covalently crosslinked triple helices that have applications as biomaterials, distinguishing it from athermal models of thick collagen fibrils[7,26-28]. Any such model must first be validated against a target class of material. We have previously characterized hydrogels synthesized *via* reaction of type-I collagen triple helices hydrogels covalently crosslinked with two diacids, 1,4-phenylenediacetic acid (4Ph), and adipic acid (AA), in terms of their hydration, biocompatibility, and bulk elastic moduli[2,30]. The hydrogels were found to display tunable mechanical and swelling properties depending on the degree of crosslinking at the molecular level. However, a systematic parameter sweep to determine optimal, application-dependent mechanical properties is experimentally challenging, hence the need for predictive pre-screening. The selection of the mathematical model was guided by the properties of individual collagen triple helices, whose force-extension curves are well described by the wormlike chain (Kratky-Porod) model in which correlations in backbone orientation decay under thermal fluctuations over the persistence length $\ell_p$ [31,32,34]. Values of $\ell_p$ quoted in the literature for collagen triple helices vary depending on the measurement method and solvent conditions used, as summarized in Pritchard *et al.*[25], but in all cases was found to be similar to or smaller than the triple helix length $L \approx 300\ nm$, which means they are best classified as semi-flexible[25,35]. Note that the mechanical properties for both flexible and semi-flexible polymers are entropic (*i.e.* thermal) in origin[35], thus thermal effects are required for realistic modeling. Crosslinked semi-flexible polymer network models have been widely studied in the context of the cellular cytoskeleton[25,35], making this a suitable starting point for representing covalently crosslinked collagen triple helices. Early semi-flexible polymer modeling extended flexible polymer theory[36,37] in deriving closed form equations under restrictive assumptions regarding network microstructure and local deformation modes[38-43]. Broader applicability can be achieved by relaxing these assumptions and incorporating additional dynamical mechanisms[44-46], although the increased complexity typically requires computational solution.

As a departure from the protocols typically employed to characterize scaffold mechanics, our research strategy is here to employ a shear protocol as the primary point of correspondence between model and experiment. This aids comparison to existing semi-flexible polymer network theory where shear is extensively used, principally because it is a volume-preserving mode that does not invoke solvent incompressibility and associated complexities[25,35]. In addition, the linear (*i.e.* small strain) response is usually validated first, before attempting to tackle the more complex non-linear regime for which many more mechanisms may need to be considered[34], and the same approach is adopted here. The material behavior of complex materials such as hydrogels can simultaneously exhibit elastic and viscous properties depending on the rate of deformation and network architecture, *i.e.* they are viscoelastic[47], and this can be quantified by employing a rheological frequency sweep under oscillatory shear. This frequency range can be used to discriminate between proposed models and therefore serve a useful purpose aiming to predict the rheological properties of hydrogel networks based on covalently crosslinked collagen triple helices experimentally synthesized in this study. Collagen suspensions were experimentally prepared in diluted acidic conditions, at which state fibrillogenesis is hindered and collagen molecules are mostly present as triple helices. Crosslinking reaction was carried at varied concentration of collagen triple helices with either aromatic, i.e. 4Ph, or aliphatic, i.e. adipic acid (AA), diacids, yielding collagen hydrogels with varied network architecture.

## 2. Materials and Methods

### 2.1 Experiments

#### 2.1.1 Synthesis of covalently-crosslinked collagen hydrogels

Three collagen solutions (0.4, 0.8 and 1.2 wt.-%) were prepared by stirring in-house isolated type I collagen in 17.4 mM acetic acid (Sigma Aldrich) at 10 °C, following which the solutions were cast into a 24-well plate (1 g solution per well). Adipic acid (AA, Sigma Aldrich) and 1,4-phenylenediacetic acid (4Ph, Sigma Aldrich) were selected as aliphatic and aromatic diacid-based crosslinkers, respectively, and activated by adapting a previous protocol[12] (Scheme 1). Briefly, AA and 4Ph were stirred in sodium phosphate buffer (0.1 M, pH 7.4, 500 µL) according to selected crosslinker molar ratios with respect to collagen lysines ($[COOH][Lys]^{-1}$: 0.5, 1 or 1.5) present in 1 g collagen solution. Equimolar amounts of 1-Ethyl-3-(3-dimethylaminopropyl)carbodiimide (EDC, Alfa Aesar) and N-hydroxysuccinimide (NHS, VWR) were added with selected molar ratios with respect to the molar content of crosslinker carboxylic functions ($[EDC][COOH]^{-1}$: 6). The solution was stirred (0 °C, 1 h) to enable NHS-mediated diacid activation, following which β-mercaptoethanol (Sigma Aldrich) ($[β-ME]=[EDC]$) was added in order to quench EDC. NHS-activated diacid solution was mixed with aforementioned collagen solution (final concentration: 0.3, 0.5 or 0.8 wt.-% collagen). Reacting mixtures were incubated overnight under gentle shaking at room temperature, resulting in complete gel formation following 24 hours. Collagen hydrogels were washed with distilled water and dehydrated in aqueous solutions of increasing ethanol concentration (0-100 wt.-% EtOH).

**2.1.2 Characterization of collagen hydrogels**

TNBS colorimetric assays were carried out in order to quantify the crosslink degree ($C$) in resulting covalent networks, as previously reported[2,12]. The rheological properties of water-equilibrated hydrogels were measured using a AR1500ex Rheometer (TA Instruments, Crawley, UK) equipped with a 60 mm plate. Hydrogels (Ø~10 mm; $h$~6 mm) were loaded onto the stage and the upper plate lowered until a 3 mm gap was reached. Loaded samples were equilibrated to 25 °C for several minutes before the measurement. A solvent trap was fitted on top of the samples to minimize solvent evaporation. Frequency sweeps were carried out with a 0.5% strain amplitude (in the linear viscoelastic region) and two replicas were used for each hydrogel group. The complex modulus $G^*(f) = G'(f) + iG''(f)$, whose real and imaginary components are the storage $G'(f)$ and loss $G''(f)$ moduli respectively, were recorded as a function of frequency $f$ [47].

**2.2 Computational model**

**2.2.1 Model components**

The mathematical model (solved computationally) was derived from a numerical scheme that has previously been applied to actin-myosin mixtures[48-50]. Each collagen triple helix was represented by a string of $M$ beads connected by Hookean springs of natural length $b$, giving a total molecular length $L = Mb$. The natural length $b$ was chosen as the base length in the simulations to which all other lengths are compared (see Section 2.2.3 below). Angular springs were included that energetically penalize deviations from straight configurations, with a controllable stiffness that can be directly related to the molecule's bending modulus $\kappa$, or equivalently its persistence length $\ell_p = \kappa/k_B T$ with $T$ the absolute temperature and $k_B$ Boltzmann's constant[36,37]; see Fig. 1(a).

In the model, crosslinker molecules were only explicitly represented when attached to two collagen molecules, when they take the form of a Hookean spring with controllable natural length and stiffness. Crosslinks are created at a rate $k^{attach}$ when two beads (not currently crosslinked) are within a range $r^{attach}$, as shown in Fig. 1(b). The attachment rate implicitly combines the reaction rates for both ends of the crosslinker to attach (as in the case of ideal covalent networks with no network defect, *e.g.* grafting) and the concentration of crosslinker in solution. We assume that this attachment rate is constant in time, corresponding to a reservoir of crosslinkers that is not substantially depleted during gelation; and uniform in space, which is valid given the free diffusion of crosslinker in solution for the scales of interest. Detachment of both heads simultaneously is also allowed at a low rate $k^{detach} \ll k^{attach}$, but this is suppressed (along with attachment) after the shear has been applied (see below).

### 2.2.2 Simulation protocol

Simulations using bespoke code were started from a configuration of randomly positioned and oriented collagen molecules at the required concentration[48,49,50]. Bead positions were updated according to Brownian dynamics, corresponding to the overdamped regime where solvent friction dominates bead inertia[51], at constant system volume. Bead overlap was avoided by including repulsive forces between nearby beads not connected by intra-molecular springs. For numerical convenience this force was taken to be a repulsive Lennard-Jones potential[51] with a range $d^{rep} = 2^{1/6} b$ with $b$ the natural length between the collagen monomers. Periodic boundary conditions were assumed along each axis and the time step corresponded to 5 ns. A snapshot is shown in Fig. 2 alongside schematics of collagen crosslinking, and a movie is provided in the Supporting Information.

After a predetermined time (the crosslinking time; see below) for the network to form and coarsen, crosslinker attachment and detachment was halted (to ensure no evolution of the network during measurement of the shear modulus) and a 5% shear strain applied *via* Lees-Edwards boundary conditions[51]. Note that this strain was larger than that used for the experiments (see Sec. 2.1.2), which was necessary to achieve a reasonable signal-to-noise ratio; however, linear response was independently confirmed. Note also that equilibrium was *not* achieved prior to the application of shear, so the crosslinking time is a control parameter. The corresponding shear stress averaged over a further time period was then used to calculate the shear modulus $G_0$. The network metrics in Fig. 1(c) were calculated just prior to applying the shear, by first identifying network nodes when collagen bundles branch or merge, and then which nodes were directly linked. Network nodes were identified as regions with high collagen density but low mean orientation of collagen molecules, and insensitivity of measured quantities to the thresholds implicit in these criteria was checked for. The network quantities are the coordination number $z$, defined as the mean number of nodes connected to any given node; the mean distance between nodes $\ell_e$; and the mean number of collagen molecules in a bundle $n_f$. Note that 'bundles' here refers to laterally crosslinked collagen triple helices, not the α chains within a helix.

The fraction of possible attachment points occupied by crosslinker, $C$, was estimated as the number of attached crosslinks anywhere in the system, $N^{cr}$, divided by the total number of beads $N^{bead}$ and the number of other molecules in a bundle $n_f - 1$, *i.e.* $C = N^{cr}/(N^{bead}(n_f - 1))$.

This calculation combines a factor of 2 (the number of attachments per crosslinker) and a factor of 0.5 (since a bead on one molecule can be attached to two offset beads on each adjacent molecule).

**2.2.3 Relationship between simulation and experimental parameters**

The simulation parameters are mapped to experimental scales by identifying and matching characteristic length, time and stress scales. The chosen length scale was that of a single collagen molecule, so $L = Mb$ corresponds to 300 nm[5]. For the time scale, the rotational diffusion time for a single rigid molecule to rotate under thermal fluctuations was used[37], and measured quantities were insensitive to time steps of around 5 ns and shorter. The total crosslinking time achieved with available computation resources corresponded to just 20 ms. To compensate for this short time, the crosslinker attachment rate was raised so that networks formed within the simulation window. To compare shear stresses, the dimensionless quantity $\sigma L^4/\kappa$ in both the simulations and in experiments was equated, where σ is the shear stress and $\kappa$ is the bending rigidity of a collagen molecule. Key simulation parameters and their corresponding real values are given in Table 1.

**3 Results and Discussion**

**3.1 Hydrogel rheological response at varied frequency**

The storage and loss moduli $G'(f)$ and $G''(f)$ for gels constructed from each crosslinker are given in Fig. 3, and demonstrate similar features. Over a frequency range 0.01-10 Hz, both moduli were flat or only weakly dependent on frequency, with $G'(f) \gg G''(f)$, demonstrating an elastic plateau. Qualitatively similar viscoelastic spectra have been observed in both self-assembled peptide[52-56] and photo-activated collagen hydrogels[57], block copolymer hydrogels[58], ionically-crosslinked intermediate filament networks[59] and *P. aeruginosa* biofilms[60], but not in crosslinked filamentous actin networks, where instead $G''(f)$ decays at low frequencies[35,61]. Indeed, since $G''(f)$ must generally approach zero as $f \to 0$ [48], all of these spectra must decay at sufficiently low frequencies, but the corresponding relaxation times may be below those accessible to the rheometer. The lack of an observed decay in $G''(f)$ can also be caused by a lack of time-translational invariance[62], arising when very slow processes extending at least as long as the experimental time frame are present.

There is therefore a broad frequency regime over which the response is dominated by a nearly-constant elasticity $G'(f = 0.01 Hz)$, and this is the quantity we focus on below, comparing it to the $G_0$ extracted from simulations. It was also evident from the figure that both moduli sharply increased for frequencies greater than 10Hz, approximately as a power law with an exponent inconsistent with existing predictions for semiflexible polymers[35,43,63-65], for which we currently have no explanation.

**3.2 Effect of the network architecture on rheological properties**

It is apparent from Fig. 3 that gels formed with AA crosslinker have a higher plateau modulus than those formed with 4Ph. In principle, the presence of the benzene ring in the 4Ph segments may promote the formation of additional and reversible $\pi - \pi$ stacking interactions, potentially resulting in increased hydrogel storage modulus, as investigated in the case of 4-vinylbenzylated collagen hydrogels[57]. In this study, however, the introduction of a covalent crosslinking segment between two collagen triple helices proceeds *via* a nucleophilic addition/elimination reaction mechanism between the NHS-activated carboxylic terminations of the same diacid molecule (either 4Ph or AA) and two primary amino functions of collagen. Given the one-step mechanism of this crosslinking reaction, the higher storage modulus of AA- compared to 4Ph-based system can be explained in terms of the increased flexibility of the aliphatic with respect to the aromatic, bulky, diacid. This enhanced flexibility feasibly increases the range of local collagen configurations enabling reaction of both carboxylic terminations of the same crosslinker molecule with primary amino groups of collagen. This will inevitably result in the formation of collagen networks with increased crosslink density (as confirmed *via* TNBS assay, Fig. 7 (top), discussed below) and elastic modulus, whilst the presence of network defects, *e.g.* grafted *vs.* crosslinked chains, will be minimized.

To probe the plausibility of the hypothesis that increased crosslinker flexibility promotes crosslinking, the model parameter for the crosslinker attachment range $r^{attach}$ was systematically varied with all other parameters kept fixed, and the plateau modulus $G_0$ measured. $r^{attach}$ was chosen as it is the physical separation within which two free collagen triple helix beads can undergo a crosslinking reaction (Sec 2.2.1), so variations in $r^{attach}$ are inherently correlated with the number of possible configurations through which covalent crosslinks are formed. Given the coarse-grained representation of the crosslinkers in the model (*i.e.* as Hookean springs), this is the most natural way to implement a controllable crosslinking propensity; in reality, $r^{attach}$ will depend on many microscopic parameters including the type of chemical bonds along the crosslinker contour and the solvent temperature. It is clear from Fig. 4 that $G_0$ does indeed increase with $r^{attach}$ (here measured relative to the bead diameter $d^{rep}$), confirming this hypothesis is plausible.

The accessibility of the model can be exploited to generate insight into this trend. As shown in Fig. 4, increasing $r^{attach}$ simultaneously alters multiple network metrics that have competing effects on the overall stiffness: The network connectivity, quantified by the coordination number $z$ ($R^2 \sim 0.45$), and the inter-node distance, $\ell_e$ ($R^2 \sim 0.27$), show only small variations, while the degree of bundling quantified by $n_f$ strongly increases ($R^2 \sim 0.98$), reflecting the increased number of configurations compatible with crosslinking. The latter trend is observed to be correlated with an increased network stiffness ($R^2 \sim 0.60$), and it is clear that bundling of triple helices controls the overall increase in stiffness in this data set.

### 3.3 Effect of diacid content on rheological properties

Fig. 5 describes the variation of the complex modulus in hydrogels obtained by crosslinking solutions with increased content of diacid. An experimental observation was that an increase in the concentration of diacid relative to that of the primary amino groups of collagen did not lead to the formation of hydrogels with increased network stiffness, whilst a non-significant decrease of rheological properties was observed (Fig. 5). In AA-crosslinked hydrogels there is no significant change in the shear modulus when increasing the 4Ph/Lys ratio by a factor of 3, whilst 4Ph-crosslinked hydrogels display a discernible downward trend over the same range.

To investigate this experimental trend *in silico*, we varied the model parameter that most closely corresponded to the crosslinker ratio, *i.e.* the attachment rate $k^{attach}$, which is proportional to the concentration of crosslinker (as opposed to $r^{attach}$, which relates to the physical properties of a single diacid; Sec. 2.2.1). Systematically varying this parameter reveals that $G_0$ weakly increases with $k^{attach}$, with a low $R^2 \approx 0.5$, as shown in Fig. 6. Plotting the various network metrics again reveals competing effects, with the network connectivity $z$ markedly ($R^2 \sim 0.84$) increasing (as expected), whilst both the degree of bundling $n_f$ ($R^2 \sim 0.63$) and the mean inter-node distance $\ell_e$ ($R^2 \sim 0.81$) decreasing, with the latter at a faster rate. Overall, the variation of the network stiffness was weaker ($R^2 \sim 0.50$) than for varying $r^{attach}$, but both exhibited positive correlations despite contrary trends in bundling, network connectivity and inter-node separations evident from comparing Figs. 4 and 6. This highlights a complex, multifactorial relationship between network microstructure and bulk hydrogel response.

The stiffness trends observed computationally are in line with the experimental observations if we consider the role of grafting. The slight variation in network stiffness observed in hydrogels prepared with increased crosslinker ratio may be related to the occurrence of network defects, *e.g.* grafted rather than crosslinked, collagen triple helices, originating from the increased probability of reaction of primary amino terminations of collagen with randomly-positioned carboxylic groups. Such grafting would result in a decrease in the extent of collagen bundling, and whereas this is indeed one consequence of increasing $k^{attach}$ as evident in Fig. 6, the effect was not strong enough to lower the predicted net stiffness. Explicitly incorporating grafting into the rules of the mathematical model may further decrease the degree of bundling and produce the same stiffness trend as in the experiments, suggesting this mechanism can increase model validity and should be considered in subsequent enhanced versions.

The relative speed of model data acquisition means that we can easily extend the range and number of data points sampled. Not shown in Fig. 5 is the limit $k^{attach} \to 0$, for which the network stiffness sharply approaches zero. Without crosslinking the hydrogel would not form, so the same limit would be observed. For fixed volume systems such as the simulations, however, this is not a trivial observation: Even without crosslinking, polymeric materials can become entangled, resulting in a finite elasticity extending down to very low frequencies that may exceed the range of the rheometer[37,38]. Entanglement occurs in polymer solutions at concentrations that can be very low for rigid molecules[40]. A vanishing $G_0$ in the no-crosslinker limit is therefore a qualitative point of correspondence between the model and the experiments.

**3.4 Tunability of crosslink density**

The fraction of attachment points (experimentally resulting in the consumption of free amino terminations and formation of amide bonds) occupied by crosslinker as quantified by both TNBS assay, and the equivalent quantity from the simulations (defined as the fraction of potential attachment points that are occupied; see Sec. 2.2.2), is presented in Fig. 7. Increasing crosslinker availability, achieved by controlling the crosslinker ratio in experiments and $k^{attach}$ in the model, results in a sublinear increase in the fraction of attached crosslinker $C$. In addition, the numerical values of $C$ for the two methods are comparable over the ranges considered, providing some support to the expedient of enhancing the attachment rate to ameliorate the short simulation times achieved.

Varying the collagen concentration in the crosslinking mixture with a fixed crosslinker ratio results in comparable values but with opposing trends, with the experimentally-determined $C$ weakly decreasing while the model prediction clearly increases. The most likely explanation is that the increased collagen concentration leads to steric hindrance of the lysines, reducing the fraction exposed for reaction. Since there is no such selectivity in the model, which permits crosslinker molecules to attach irrespective of collagen orientation, steric hindrance is absent and the contrary trend is observed. As in Sec. 3.3, the partial correspondence between model and experiments has aided the identification of a putative mechanism underlying the experimental trend by their absence from the current model.

**3.5 Effect of collagen triple helix concentration on rheological properties**

There exist many semiflexible polymer network models that differ in their assumptions regarding the properties of individual fibres and how they interconnect into the network[35,38,40-42]. All predict a power law increase in the elastic modulus as the polymer concentration increases, but with differing exponents; thus determining this exponent experimentally can suggest representative models. This has been attempted for peptide gels, for instance[56,66,67]. Note that these models make many more assumptions than the detailed computational model considered here, but this allows analytical expressions to be derived without the need for intensive numerical integration.

The increase in shear modulus in collagen hydrogels prepared from crosslinking mixtures with increased collagen triple helix concentration was observed both by the experiments and the simulations and is shown in Fig. 8. Both increase approximately as a power law, but with differing exponents. Fitting the simulation data gave an exponent of 1.67±0.07, consistent with one model prediction[42] of 5/3. However, the experimental data exhibits a far greater exponent exceeding 3, outside of the range of existing theoretical predictions, although comparable to one class of peptide gels at pH=4 [47]. We note that networks of physiological collagen fibrils (where thermal fluctuations are not relevant) exhibit an exponent closer to 2 [26,28,68].

By combining existing models, it is possible to derive an exponent of 4 that is consistent with the experimental data. This is based on three assumptions: (a) Network nodes are connected by tightly-coupled bundles consisting of crosslinked collagen triple helices similar to Fig. 1(c) [63,64]; (b) The network microstructure corresponds to rod-like bundles[39,39]; and (c) The force-extension relation for changes in connected node separation is purely entropic. This latter assumption applies as long as the degree of bundling is low, valid for the hydrogels under consideration as the crosslinking was carried out in diluted acid conditions that do not favor fibrillogenesis (which would lead to extensive bundling dominated by enthalpic contributions[7,26-28]). With these assumptions, it is possible to derive the prediction

$$G_0 = \frac{3\pi^3}{4} n^4 L^4 \kappa \ell_p b^3 n_f^{3/2} \qquad (1)$$

where $n$ is the number density of collagen molecules, $\kappa$ is the bending modulus of a single collagen molecule, $\ell_p$ is its persistence length, $b$ is the lateral separation of collagen within a bundle, and $n_f$ is the number of molecules in a bundle as before. Details of this derivation are given in the Supporting Information.

Fitting the experimental data to the form $G_0 \propto (concentration)^4$ as predicted by equation (1) gives a reasonable fit to both the AA and 4Ph crosslinked gels, as shown in Fig. 8. Furthermore, all quantities in equation (1) are either known or can be estimated as already described in Section 2.2, with the exception of $n_f$. Estimates for $n_f$ can therefore be determined from the fit, and they are consistent with the expectation of limited bundling for these hydrogels, giving $n_f \approx 4.5$ for AA and $n_f \approx 3.5$ for 4Ph. It is also shown in the Supporting Information that assumption (b) is valid for these bundle sizes, confirming the self-consistency of this calculation. The assumptions behind equation (1) were therefore consistent with the experimental data; however, it was too basic to permit fitting to the other quantities discussed above, for which we used the higher-fidelity simulation model.

As a demonstration of the model's predictive potential in its current form, the shear modulus systematically varying both the collagen concentration and the concentration of crosslinker is provided in Fig. 9. From inspection it was immediately clear that the stiffest networks were to be found in the region of high collagen concentration, but only weakly depended on the crosslinker ratio once a threshold value had been exceeded. This figure contains 51 distinct combinations of control parameters, each averaged over 20 repeats, and required around 4 weeks of local cluster time; it would be completely unfeasible to perform corresponding *in vitro* assays. Plots such as this could be used to guide the selection of control parameters for subsequent *in vitro* investigation, reducing the number of lengthy and costly experiments required to determine the optimal parameter combination for real materials.

## 4. Conclusions

There are significant potential benefits to providing a predictive capability for the design of novel hydrogels with tailored mechanical properties. The model that has been detailed here and critically compared to rheological data was found to reproduce key experimental trends, with quantitative correspondence. In addition to playing a predictive role, microscopic models such as this provide mechanistic insight into the changes in network microstructure underlying the variations in bulk mechanics, such as the variation with crosslinker flexibility discussed in Sec. 3.2. In addition, a simple model for the increase in stiffness with collagen concentration fitted well to the data, allowing a network metric (the degree of bundling) to be extracted from rheology data without recourse to scattering or electron microscopy.

Although the correspondence between the model and experiments is promising, there remain deviations that will need to be corrected before full validation can be claimed. Some possible reasons for these and means to address them, namely the inclusion of explicit grafting and steric hindrance between lysines, have already been discussed in Sec. 3.3 and 3.4 and could be addressed by explicitly representing diffusing crosslinkers in solution, each with a pair of reactive sites. An additional cause may be the limited simulated times achieved, of the order of just tens of milliseconds of crosslinking time, in contrast to the hour scale needed to observe experimental gelation; this was ameliorated by artificially increasing the crosslinker attachment rate. It is possible to drastically extend simulated times by using implicit methods[64] (rather than the explicit method used here), or extending the code's shared memory parallelism to exploit distributed memory and take advantage of more of the cluster's computational resources[69]. Both enhancements require significant development time and are beyond the scope of this investigation, but can be considered now that the strengths and deficiencies of the simpler model described herein have been identified.

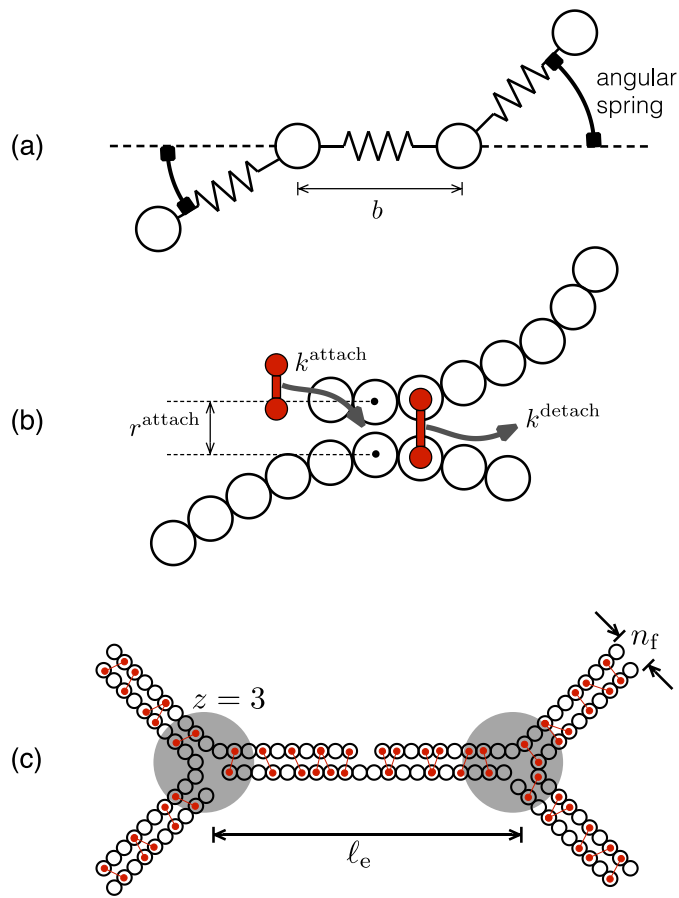

**Figure 1.** Schematic representation of key model components. (a) Bead-spring representation of collagen triple helices showing longitudinal springs controlling molecule length fluctuations (zig-zag lines), and angular springs controlling bending rigidity (thick lines with square heads). (b) Crosslinkers attach with a rate $k^{attach}$ when two bead centers are within a distance $r^{attach}$, and detach at a rate $k^{detach}$. Springs not shown for clarity. (c) Example of network structure, showing two nodes (highlighted with grey discs) with a coordination number $z = 3$. The distance between nodes is $\ell_e$, and all bundles consist of $n_f = 2$ molecules. Online version in color.

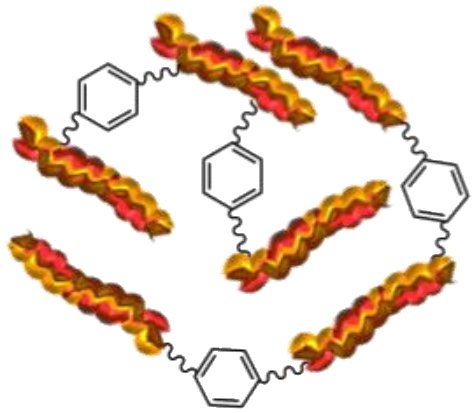 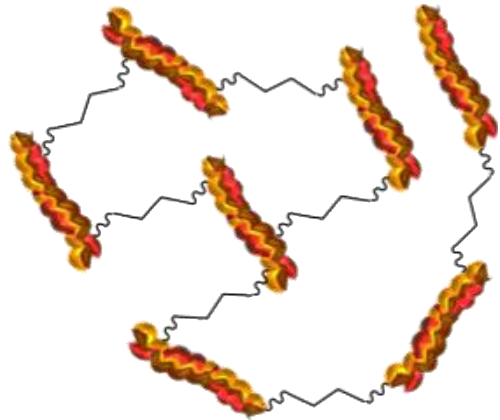

(a)          (b)

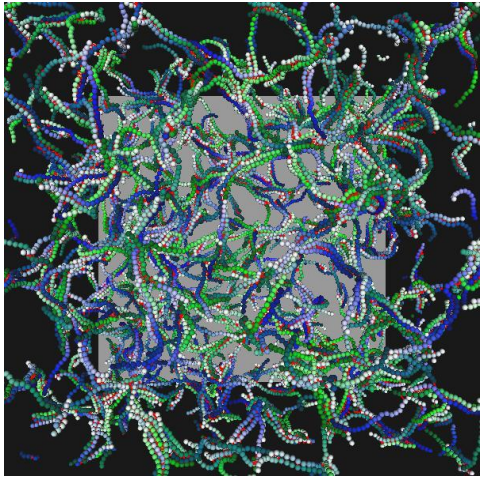 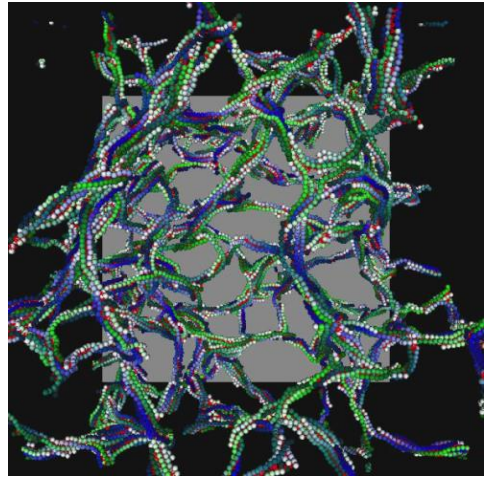

(c)          (d)

**Figure 2.** (a) Schematic representation of the covalent network architecture in collagen hydrogels crosslinked with 4Ph. (b) The same with AA as the crosslinker. The lower two panels show snapshots of the computational model at early (c) and late (d) times for the same run. The collagen molecules are strings of M=20 beads (coloring and shading purely to aid visualization), connected by short straight crosslinkers. Periodic boundary conditions are assumed along each axis. Online version in color.

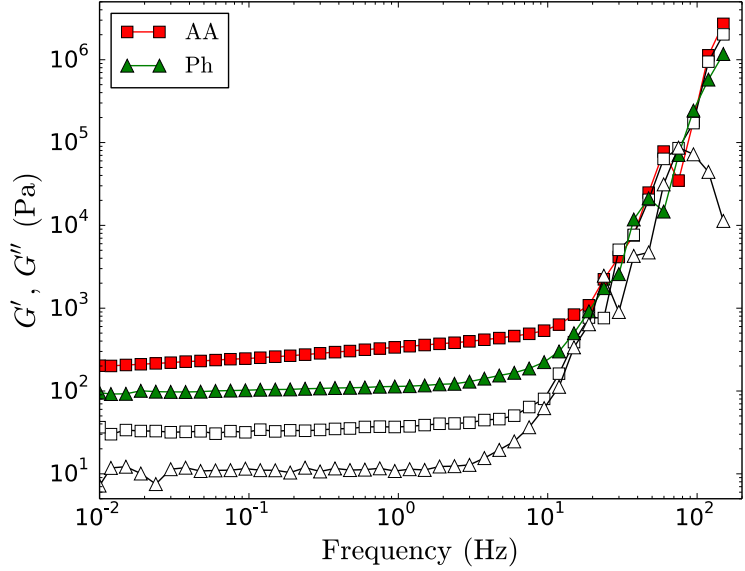

**Figure 3.** Linear viscoelastic spectra of collagen gels formed by two types of crosslinker, showing $G'(f)$ (filled symbols) and $G''(f)$ (open symbols). The crosslinking molecules were AA (squares) and 4Ph (triangles). In both cases the collagen concentration (as measured in the crosslinking mixture) was 0.5 wt% and the crosslinker ratio between carboxylic and lysine terminations was 1. Online version in color.

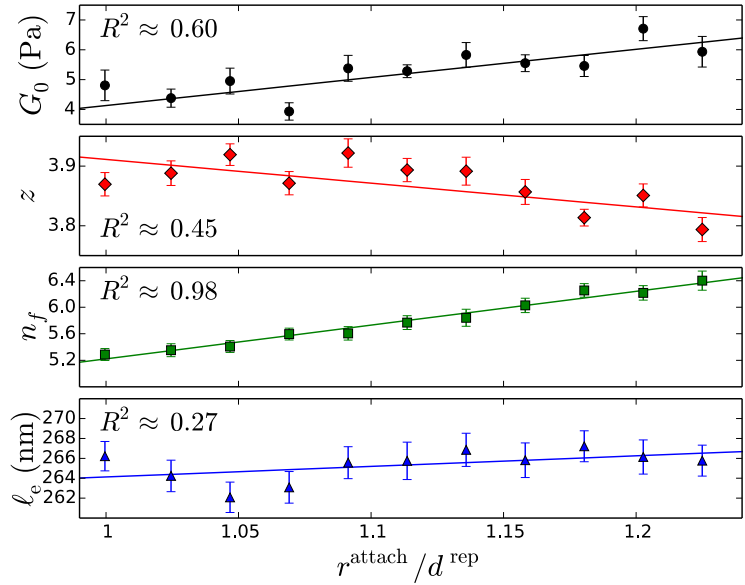

**Figure 4.** Simulation results varying the attachment range of crosslinkers $r^{attach}$ normalized to the range of repulsion between two beads $d^{rep}$. Shown are the zero-frequency shear modulus $G_0$, coordination number $z$, bundle thickness in terms of the number of triple helices $n_f$, and the distance between network nodes $\ell_e$; see Fig. 1(c). Linear lines of best fit are given with their corresponding coefficients of determination $R^2$. Here and below, error bars are standard errors (sample size n=20). Online version in color.

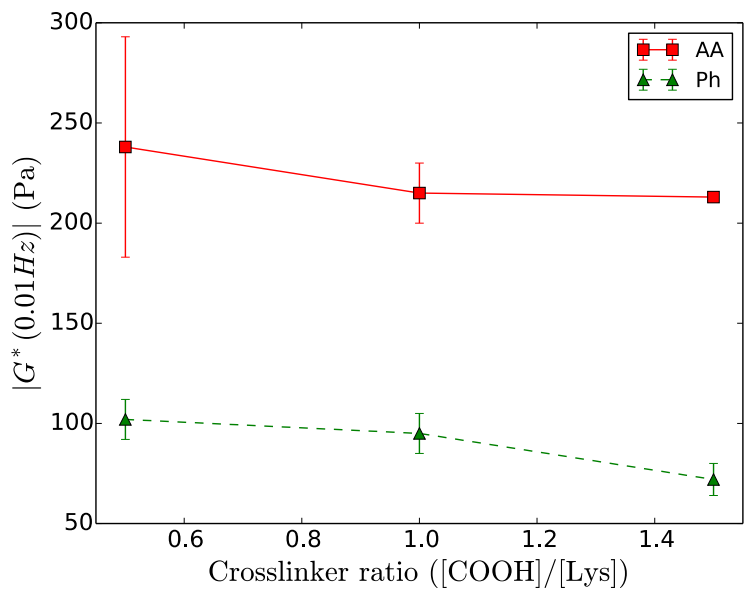

**Figure 5.** Experimental plateau moduli for gels constructed from each crosslinker, versus the ratio of crosslinker to potential attachment sites. In all cases the collagen concentration in the crosslinking mixture was 0.5 wt%. Lines are to guide the eye. Online version in color.

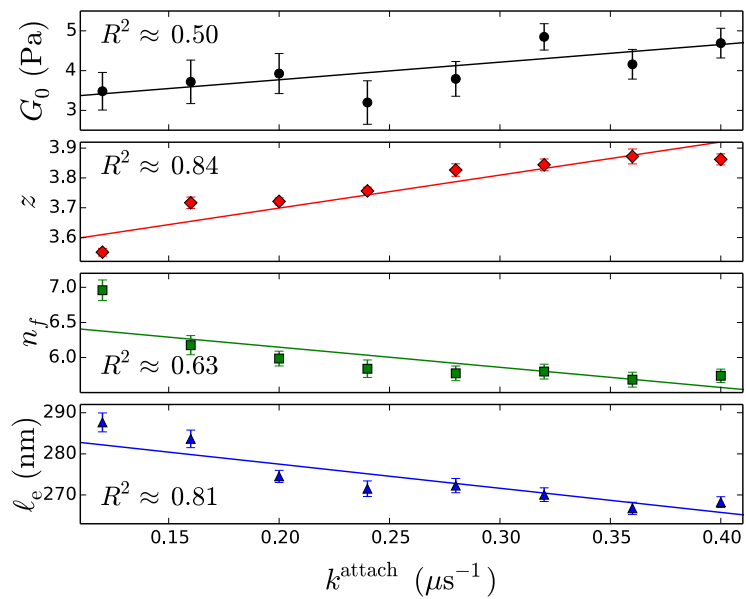

**Figure 6.** Simulation results varying the attachment rate $k^{attach}$, which is proportional to the concentration of crosslinker and hence the crosslinker ratio. Plotted quantities are the shear modulus $G_0$, the network connectivity $z$, the degree of bundling $n_f$ and the separation between network nodes $\ell_e$. Online version in color.

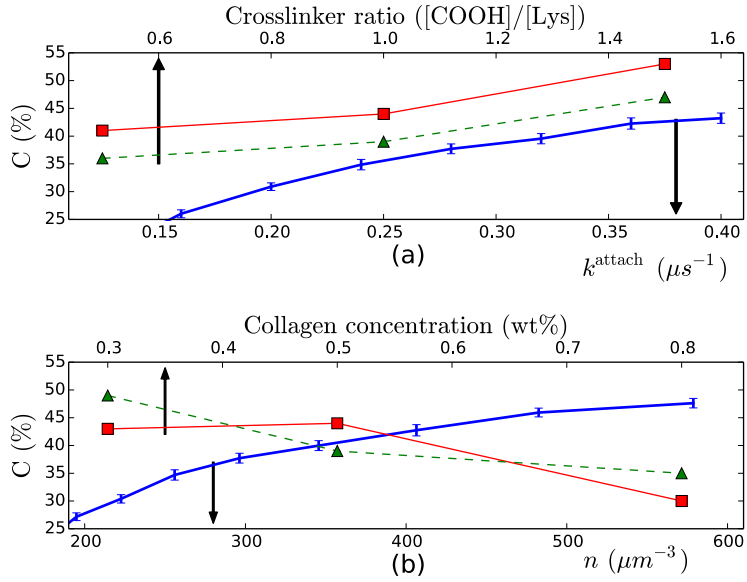

**Figure 7.** The fraction of attached crosslinks $C$, varying (a) the crosslinker availability with a fixed collagen concentration in the crosslinking mixture, and (b) the collagen concentration in the crosslinking mixture at fixed crosslinker availability. For both plots, the upper horizontal axis (symbols, thin lines) corresponds to experimental data, and the lower axis (thick lines) to the model. The experimental symbols correspond to AA (squares) and 4Ph (triangles). The two axis scales were aligned by applying a linear scaling: $k^{attach}$ was 0.25 μs$^{-1}$ times the crosslinker ratio, and the number density $n$ was 700 μm$^{-3}$ times the collagen concentration in wt%. Online version in color.

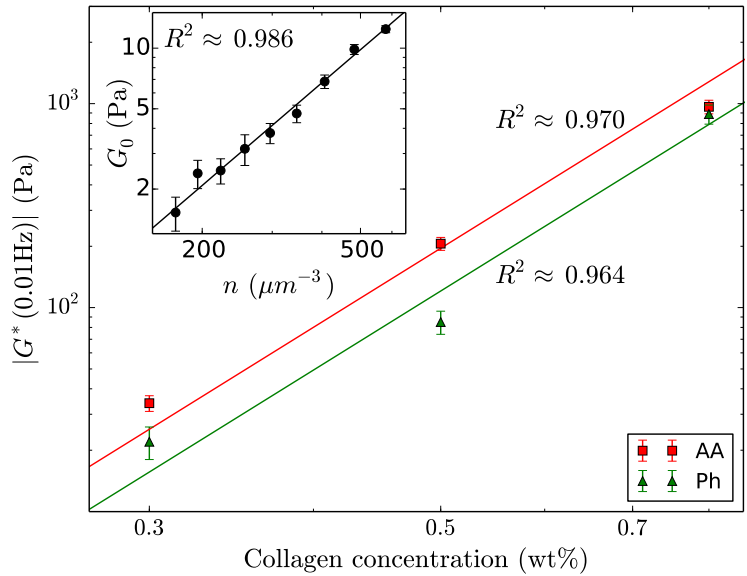

**Figure 8.** Elasticity versus concentration of collagen molecules for the two crosslinkers. Straight lines gives best fits to $G \propto (concentration)^4$ and their $R^2$. (Inset) Results from the simulations showing the shear modulus as a function of the number concentration of collagen molecules. The straight line shows best fit to $G_0 \propto n^a$ with the exponent $a$ also a fit parameter. Online version in color.

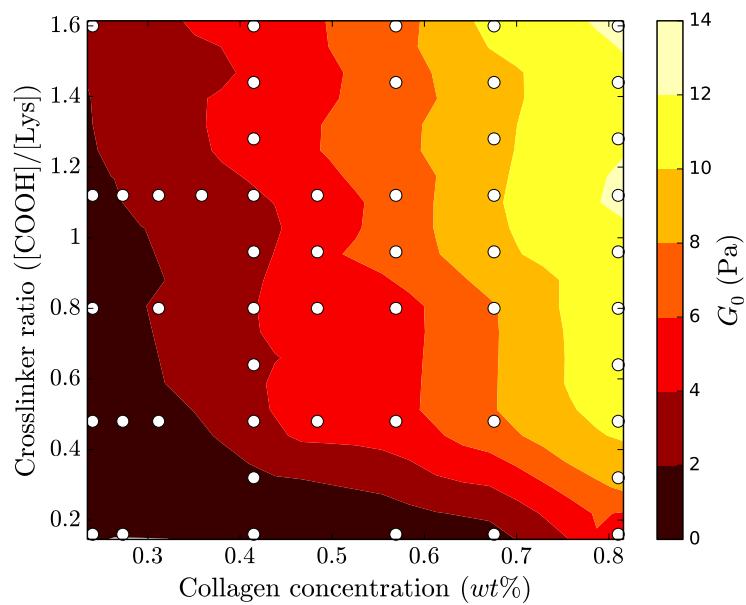

**Figure 9.** Model predictions for the network stiffness $G_0$ (calibration bar on the right) varying both the collagen concentration and the crosslinker ratio, using the same scaling factors as Fig. 7. White discs represent actual data points averaged over 20 repeats, and linear interpolation was used elsewhere. Online version in color.

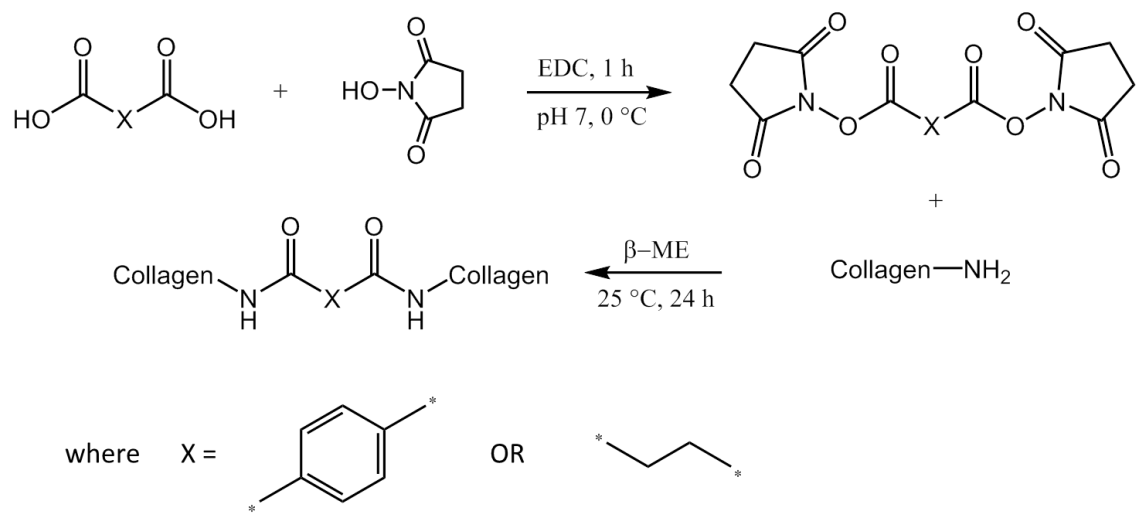

**Scheme 1:** Synthesis of covalently crosslinked collagen hydrogels. A diacid crosslinker (either AA or 4Ph) is NHS-activated in the presence of EDC. Following β-ME quenching of EDC, collagen is reacted with the activated diacid, leading to the formation of a covalent hydrogel network.

**Table 1.** Key simulation parameters and their equivalent experimental values, with references given where relevant.

| Parameter | Symbol | Simulation value | Dimensional equivalent | Ref. |
|---|---|---|---|---|
| **Rotational diffusion** | $D_r^{-1}$ | $2 \times 10^3$ | 5 ms | 37 |
| **Molecule length** | $L$ | 20 | 300 nm | 5 |
| **Persistence length** | $\ell_p$ | 3.3 | 50 nm | 25 |
| **Attach rate** | $k^{attach}$ | 0.7 | 0.28 µs$^{-1}$ | - |
| **Detach rate** | $k^{detach}$ | $10^{-2}$ | $4 \times 10^{-3}$ µs$^{-1}$ | - |
| **Attach range** | $r^{attach}$ | $2^{1/6}$ | 17 nm | - |
| **Bead repulsion range** | $d^{rep}$ | $2^{1/6}$ | 17 nm | - |
| **Crosslinker length** | $\ell_0^{cr}$ | $2^{1/6}$ | 17 nm | - |
| **Crosslinker stiffness** | µ | 5 | $3 \times 10^{-2}\ pN/nm$ | - |

**Supporting Information**. The Supporting Information is available free of charge on the ACS Publications website at DOI: 10.1021/acsbiomaterials.6b00115.

**Corresponding Author**

*E-mail: d.head@leeds.ac.uk.

**Author Contributions**

DH designed and implemented the numerical model, carried out the simulations, and participated in the data analysis; GT designed the experimental study, performed the experiments, and participated in the data analysis; DJW and SJR participated in the design of the experimental study. All authors helped in the preparation of the manuscript and gave final approval for publication.

**Funding Sources**

DH was funded by the Biomedical Health Research Centre, University of Leeds. GT and SJR wish to thank The Clothworkers' Centre for Textile Materials Innovation for Healthcare and the Clothworkers' Foundation for financial support.